\documentclass[preprintnumbers,showpacs,amsmath,amssymb,floatfix,prd,onecolumn,superscriptaddress,nofootinbib,12pt]{revtex4}
\usepackage{amssymb}
\usepackage{graphicx}
\usepackage{epsfig}
\usepackage{bm}
\usepackage{mathrsfs}
\usepackage{amsfonts}
\usepackage{epstopdf}
\usepackage{bm}
\usepackage{amsthm}
\usepackage{multirow}
\usepackage{subfigure}
\usepackage{amsmath}
\usepackage{textcomp}
\usepackage{fancyhdr}
\usepackage{tipa}
\usepackage{babel}
\begin{document}
\title{Generalized Off-Shell ADT Conserved Charges in the Presence of Matter Chern-Simons Term}

\author{Hai-Feng \surname{Ding}}
\email[]{haifeng1116@qq.com}
\author{Xiang-Hua \surname{Zhai}}
\email[]{zhaixh@shnu.edu.cn}
\affiliation{Division of Mathematical and Theoretical Physics, Shanghai Normal University, 100 Guilin Road, Shanghai 200234, China}

\begin{abstract}
In this paper we elevate the formalism for off-shell Abbott-Deser-Tekin (ADT) conserved charges in the presence of arbitrary matter fields to including the internal gauge transformation, when the gauge fields are present. For this purpose, we resort to exact symmetry, and the symmetry generator is combined by diffeomorphism and internal gauge transformation. For universality, we consider an apparently non-covariant Lagrangian, which contains a matter Chern-Simons term. We show that the elevated off-shell ADT formalism is equivalent to the covariant phase space method and the Barnich-Brandt-Comp\`ere (BBC) formalism. To check the validity of our method, we explicitly compute the conserved charges of general non-extremal rotating charged G\"odel black holes in minimal five-dimensional gauged supergravity, reproducing the previously known results.
\end{abstract}

\maketitle


\maketitle

\newpage
\section{Introduction}

Symmetry plays an important role in modern physics. The well known Noether's theorem, relating symmetries to conserved charges, has become a cornerstone in modern physics and provided a deep basis to understanding the conservation laws in classical mechanics, general relativity and quantum field theory. However, Symmetry is not so straightforward to define conserved charges in an unambiguous manner in general theory of gravity. On the other hand, because of the restriction of equivalence principle, till now there is no general consensus for defining the local conservation quantities in general relativity. At most, one can define quasi-local conserved charges in finite spacetime domains (for quasi-local concepts and an extensive review of conserved charges, please refer to ref.\cite{Szabados:2004vb}). Up to now, several approaches have been proposed to compute the conserved quantities, each having their own merits and demerits.

In the well known ADM approach \cite{Arnowitt:1962hi}, the conserved charges could be calculated for asymptotically flat spacetime at the asymptotic infinity. But this approach fails for anti-de Sitter (AdS) spacetime.  Extension to asymptotic AdS geometry has been given by Abbott, Deser and Tekin (ADT)
\cite{Abbott:1981ff,Abbott:1982jh,Deser:2002rt,Deser:2002jk,Senturk:2012yi} in a covariant manner. In this  traditional ADT method, one needs to linearize the dynamical field and use the equations of motion (EOM), i.e. on-shell. As a result, the procedures become highly complicated when the higher curvature or higher derivative terms are present in the Lagrangian. Without resorting to the linearization of dynamical field, the Brown-York method is put forwarded in \cite{Brown:1992br} by introducing an appropriate counter term. This formulation is especially useful in the context of the AdS/CFT correspondence. But usually, it is difficult to find an appropriate counter term, and this method is not covariant.

Based on the Noether procedure, the covariant phase space method (CPSM) was proposed in \cite{Wald:1993nt, Iyer:1994ys, Jacobson:1993vj, Wald:1999wa} by Wald and Iyer (see \cite{Hajian:2015eha, Seraj:2016cym, Corichi:2016zac} for reviews ). In the framework of CPSM, a convenient method for calculating conserved charges associated with `exact symmetries' of black hole solutions in generally covariant gravitational theories was developed by K. Hajian, which is called solution phase space method (SPSM) \cite{Hajian:2015xlp} (see \cite{Hajian:2016kxx, Ghodrati:2016vvf, Hajian:2016iyp, Chernyavsky:2017xwm} for reviews and applications of SPSM). Along another line, general theory of conserved charges based on the cohomology principles was developed by Barnich-Brandt and Comp\`ere (BBC) \cite{Barnich:2001jy, Barnich:2003xg, Barnich:2004uw, Barnich:2007bf} (see \cite{Compere:2007az,Compere:2018aar} for reviews ). These formalisms have been extensively applied to investigate the black hole thermodynamic properties, especially used to derive the first law of black hole mechanics for generic theories of gravity. Recently, they have also been applied to investigate the holographic properties \cite{Faulkner:2013ica}.

Another interesting method to calculate the quasi-local conserved charges in a covariant theory of gravity is proposed by Kim, Kulkarni and Yi \cite{Kim:2013zha}, which generalizes the on-shell Noether potential in the ADT formalism to off-shell level. This method is based on the Lagrangian description, and so completely covariant. In technology, it uses the one parameter path integral method developed in \cite{Wald:1999wa,Barnich:2001jy, Barnich:2003xg, Barnich:2004uw, Barnich:2007bf} in solution phase space to obtain the finite charge difference. It works well for higher curvature case.  In \cite{Kim:2013cor}, the quasi-local formulation of conserved charges has been extended to a gravity theory  containing a gravitational Chern-Simons term, and it was shown that this quasi-local extension of ADT method is very effective even to an apparently non-covariant Lagrangian. Extension to generally covariant theory of gravity in the presence of arbitrary matter fields was presented in \cite{Hyun:2014sha}. The case of including asymptotic Killing vectors has also been considered in \cite{Hyun:2014kfa}. The applications of off-shell ADT formalism for various interesting geometries were presented in \cite{Gim:2014nba,Hyun:2016dvt,Kulkarni:2019nrk,Peng:2014gha,Peng:2015yjx,Peng:2016wzr} and references therein, for extensive review see \cite{Adami:2017phg}.

In \cite{Hyun:2014sha}, the quasi-local formalism for conserved charges has been extended to a theory of gravity with arbitrary matter fields, but the internal gauge transformation was not included in. In many cases, if the gauge fields are present, like in Einstein-Maxwell or Einstein-Yang-Mills theories, one needs to consider a specific gauge transformation, which will lead to corresponding conserved charges. In these cases the symmetry generators are combined by diffeomorphism and $U(1)^{(n)}$ gauge transformation. CPSM including the internal gauge freedom was considered systematically in \cite{Prabhu:2015vua} on a principle bundle. For the off-shell ADT formalism the internal gauge transformation was considered first in \cite{Setare:2016dex}. But different from \cite{Setare:2016dex} , to obtain the conserved current and potential, we resort to exact symmetries, which are well appreciated in SPSM \cite{Hajian:2015xlp, Hajian:2016kxx, Ghodrati:2016vvf, Hajian:2016iyp, Chernyavsky:2017xwm}. For universality, we consider an apparently non-covariant Lagrangian, which contains a matter Chern-Simons term. The non-covariant Lagrangian case was first provided by Tachikawa \cite{Tachikawa:2006sz} in Iyer-Wald formalism, and applied to calculate central charges in extreme Black Hole/CFT correspondence in \cite{Compere:2009dp}. For more discussion see \cite{Mast thesis}.

For quite some time, the calculation of conserved charges is a hard problem for G\"odel type black hole \cite{Gimon:2003ms, Banados:2005da, Barnich:2005kq,Wu:2007gg}, which is the exact solution of Einstein-Maxwell Lagrangian with a matter Chern-Simons term. The BBC method provides an effective way to calculate the conserved charges for this type of solution \cite{Barnich:2005kq,Banados:2005da}, which are previously obtained just via indirect method.
We show that the elevated off-shell ADT method is completely equivalent to the CPSM and also the BBC formalism. To check the validity of our method, we give a specific example for calculating the conserved charges in minimal five-dimensional gauged supergravity.

The paper is organized as follows. In Sect.2, we generalize the off-shell ADT formalism of conserved charges in the presence of arbitrary matter fields to including the internal gauge transformation, when the gauge fields are present. For universality, we consider an apparently non-covariant Lagrangian. We show that the elevated off-shell ADT formalism is identical with the CPSM and the BBC formalism. In Sect.3, we give a general model of Einstein-Maxwell-Scalar theory with cosmological constant and matter Chern-Simons term for calculating the off-shell ADT potential and charges. In Sect.4, we explicitly compute the conserved charges of general non-extremal rotating charged G\"odel black holes in minimal five-dimensional gauged supergravity. The conclusion and outlook are given in the last section.

\section{Generalized off-shell ADT current and potential}
In this section we elevate the off-shell ADT conserved charge formalism with arbitrary matter fields to including the internal gauge transformation, when the gauge fields are present. We construct the off-shell ADT current and potential, which will provide an effective way to compute conserved charges.

\subsection{Formalism}
Following \cite{Kim:2013cor,Hyun:2014sha}, we consider an action which contains a non-covariant term and with arbitrary matter fields $\psi =\left(A_{\mu }^I,\phi ^C,\cdots \right)$
\begin{equation}\label{action}
I[g,\psi ]=\frac{1}{16 \pi  G}\int d^Dx\sqrt{-g}L(g,\psi ) .
\end{equation}
For convenience, we denote the dynamic fields jointly as $\Phi =\left(g_{\mu \nu },\psi \right)=\left(g_{\mu \nu },A_{\mu }^I,\phi ^C,\cdots \right)$.
The generic variation of the Lagrangian leads to
\begin{alignat}{23}
\delta  \left(\sqrt{-g} L\right)&=\sqrt{-g} \mathcal{E}_{\Phi }\delta \Phi+\sqrt{-g}\nabla _{\mu }\Theta ^{\mu }(\delta \Phi ,\Phi )\\&=\sqrt{-g} \left(- \mathcal{E}^{\mu \nu }\delta g_{\mu \nu }+\mathcal{E}_{\psi }\delta \psi\right)+\partial_{\mu } \tilde{\Theta }^{\mu }(\delta \Phi ,\Phi ) ,
\end{alignat}
where $\mathcal{E}_{\Phi }=\left(\mathcal{E}_{\mu \nu }, \mathcal{E}_{\psi }\right)$ and $\tilde{\Theta }^{\mu }=\sqrt{-g} \Theta ^{\mu }$ denote the Euler-Lagrange expression and the surface term, respectively. And hereinafter, the tilde $\tilde{} $  over a letter denotes $\tilde {X}=\sqrt{-g}X$.

In order to introduce the off-shell current and potential, we consider a vector field $\xi =\xi ^{\mu }\partial _{\mu }$ defined over the spacetime, which generates the diffeomorphism $x^{\mu }\to x^{\mu }-\xi ^{\mu }$. In addition, if some number of $U(1)^{(n)}$ gauge fields are present, we might have gauge transformation $A^I\to A^I+d\lambda^I$ for arbitrary scalars $\lambda ^I$, under which the Lagrangian is invariant. In this case we can denote the generator by $\epsilon =(\xi ,\lambda ^I)$ , such that
\begin{equation}\label{key7}
\delta_{\epsilon }\Phi=\delta _{\xi }\Phi+\delta _{\lambda ^I}A^I .
\end{equation}
The variation of Lagrangian induced by $\epsilon$ is given by
\begin{align}
	\delta _{\epsilon }\left(\sqrt{-g} L\right)&=\sqrt{-g}   \mathcal{E}_{\Phi }\delta _{\epsilon }\Phi +\partial_{\mu } \tilde{\Theta }^{\mu }\left(  \delta _{\epsilon }\Phi,\Phi \right)    \nonumber  \\&=\sqrt{-g} \left(- \mathcal{E}^{\mu \nu }\delta _{\epsilon } g_{\mu \nu }+\mathcal{E}_{\psi }\delta _{\epsilon } \psi\right)+\partial_{\mu} \tilde{\Theta }^{\mu }\left(\delta _{\epsilon }\Phi,\Phi \right)    \nonumber  \\&=\sqrt{-g} \left(-\nabla _{\mu }\left(2 \mathcal{E}^{\mu \nu } \xi _{\nu }\right)+2 \xi _{\nu } \nabla _{\mu }\mathcal{E}^{\mu \nu }+ \mathcal{E}_{\psi }\delta _{\epsilon }\psi\right)+\partial_{\mu} \tilde{\Theta }^{\mu }\left(\delta _{\epsilon }\Phi,\Phi \right) ,
	\label{diffv1}
\end{align}
where $\delta _{\epsilon } g_{\mu \nu }=\mathcal{L}_{\xi }g_{\mu \nu } =2\nabla _{(\mu }\xi _{\nu )}$, and $\mathcal{L}_{\xi }$ is Lie derivative along the vector $\xi$.

On the other hand, the non-covariant Lagrangian transforms as \cite{Tachikawa:2006sz}
\begin{equation}\label{diffv2}
\delta _{\epsilon }\left(\sqrt{-g} L\right) =\partial_ {\mu} \left(\sqrt{-g} \xi ^{\mu} L +\tilde{\Xi }^{\mu }\right) ,
\end{equation}
where $\tilde{\Xi }^{\mu }$ denotes an additional surface term which arises from the non-covariance of Lagrangian. By equating the two expressions above, we can obtain an off-shell identity
\begin{equation}\label{identiti}
2 \xi _{\nu } \nabla _{\mu }\mathcal{E}^{\mu \nu }+   \mathcal{E}_{\psi }\delta _{\epsilon }\psi=\nabla _{\mu }\left(\mathcal{Z}^{\mu \nu } \xi _{\nu}\right) ,
\end{equation}
where we have defined
\begin{align}
	\label{33}
	\sqrt{-g} \mathcal{Z}^{\mu \nu }\xi _{\nu } \equiv \sqrt{-g}\xi ^{\mu } L +\tilde{\Xi }^{\mu }-\tilde{\Theta }^{\mu }\left(\delta _{\epsilon }\Phi,\Phi \right)+2 \sqrt{-g} \mathcal{E}^{\mu \nu }\xi _{\nu }+\partial _{\nu }\tilde{U}^{\left[\mu \nu \right]},
\end{align}
and $\tilde{U}^{\mu \nu }=\tilde{U}^{\left[\mu \nu \right]}$ is an arbitrary anti-symmetric second rank tensor, and we drop  it out in what follows since this ambiguity will not affect the final result.
Furthermore, Eq.\eqref{identiti} can be written as
\begin{equation}\label{identity0}
\nabla _{\mu }\left(2 \textbf{E}^{\mu \nu } \xi _{\nu}\right)= \mathcal{E}^{\mu \nu }\delta _{\epsilon } g_{\mu \nu }-\mathcal{E}_{\psi }\delta _{\epsilon }\psi=-\mathcal{E}_{\Phi } \delta _{\epsilon }\Phi ,
\end{equation}
where
\begin{equation}\label{key3}
\textbf{E}^{\mu \nu } \equiv \mathcal{E}^{\mu \nu }-{\frac{1}{2}}\mathcal{Z}^{\mu \nu }.
\end{equation}
If the transformation $\delta _{\epsilon }\Phi$ is an exact symmetry, i.e. $\delta _{\epsilon }\Phi=0$ (for more details and discussion we refer the readers to SPSM \cite{Hajian:2015xlp, Hajian:2016kxx, Ghodrati:2016vvf, Hajian:2016iyp, Chernyavsky:2017xwm}, and at this stage we do not restrict the field $\Phi $ to satisfy EOM), we can get
\begin{equation}
\nabla _{\mu }\left(\textbf{E}^{\mu \nu } \xi _{\nu}\right)=0 .\label{conserved1}
\end{equation}

 To obtain the off-shell conserved current, we consider the double variations \cite{Hyun:2014sha}
 \begin{equation}\label{key1}
 \delta _1 \delta _2 I\left[\Phi \right] =\frac{1}{16 \pi  G}\int d^Dx \left[\delta _1 \left(\sqrt{-g}\mathcal{E}_{\Phi }\delta _2\Phi\right)+\partial _{\mu }\left(\delta _1\tilde{\Theta }^{\mu }\left(\delta _2 \Phi ,\Phi \right)\right)\right] .
 \end{equation}
 Using the property
 \begin{equation}\label{key}
 (\delta _1 \delta _2-\delta _2 \delta _1) I\left[\Phi \right]=0 ,
 \end{equation}
 and taking one of the varations as $\delta _{\epsilon }$, we have
 \begin{equation}\label{key}
 0=\frac{1}{16 \pi  G}\int d^Dx\left[\delta _{\epsilon } \left(\sqrt{-g} \mathcal{E}_{\Phi }\delta \Phi\right)-\delta  \left(\sqrt{-g}\mathcal{E}_{\Phi }\delta _{\epsilon }\Phi\right)+\partial_{\mu} \tilde{\omega }^{\mu } \left(\delta \Phi ,\delta _{\epsilon }\Phi,\Phi \right)\right] ,
 \end{equation}
where we have used the symplectic current definition in CPSM \cite{Wald:1993nt, Iyer:1994ys, Jacobson:1993vj, Wald:1999wa}
\begin{equation}\label{symplectic current}
\tilde{\omega }^{\mu }\left(\delta  \Phi ,\delta _{\epsilon }\Phi,\Phi \right)=\delta \tilde{\Theta }^{\mu }\left(\delta _{\epsilon} \Phi,\Phi \right)-\delta _{\epsilon }\tilde{\Theta }^{\mu }(\delta \Phi ,\Phi ).
\end{equation}
Since the Euler-Lagrange expression is covariant, we can use the exact symmetry $\delta _{\epsilon }\Phi  =0$ and $\tilde{\omega }^{\mu }\left(\delta \Phi ,\delta _{\epsilon }\Phi,\Phi \right)=0$ to obtain the formula
\begin{equation}
\delta _{\epsilon } \left(\sqrt{-g} \mathcal{E}_{\Phi }\delta \Phi\right)=\delta _{\xi } \left(\sqrt{-g} \mathcal{E}_{\Phi }\delta \Phi\right)+\delta _{\lambda } \left(\sqrt{-g} \mathcal{E}_{\Phi }\delta \Phi\right)=\partial_ {\mu} \left(\xi ^{\mu } \sqrt{-g}\mathcal{E}_{\Phi }\delta \Phi\right)=0 . \label{conserved2}
\end{equation}
 Now we can introduce an off-shell ADT current for any diffeomorphism $\xi$ as in \cite{Hyun:2014sha} in the presence of arbitrary matter fields
  \begin{equation}
  \sqrt{-g} \mathcal{J}_{ADT}^{\mu }=\delta  \left(\sqrt{-g} \textbf{E}^{\mu \nu } \xi _{\nu }\right)+\frac{1}{2}   \sqrt{-g} \xi ^{\mu } \mathcal{E}_{\Phi }\delta \Phi ,
  \label{off-shell ADT current}
  \end{equation}
where we have taken $\delta\xi ^{\mu }=0$, $\delta {\lambda ^I}=0$, i.e. the generators are field independent.
From Eqs.\eqref{conserved1} and \eqref{conserved2}, it is easy to see the conservation of the Off-shell ADT current as
\begin{equation}\label{key}
	 \partial_ {\mu}\left(\sqrt{-g} \mathcal{J}_{ADT}^{\mu }\right)=0 .
\end{equation}
So we are allowed to introduce the off-shell ADT potential $\mathcal{Q}_{ADT}^{\mu \nu }$ as
\begin{equation}\label{key}
\mathcal{J}_{ADT}^{\mu }=\nabla _{\nu }\mathcal{Q}_{ADT}^{\mu \nu } .
\end{equation}
\subsection{The correspondence of off-shell ADT potential and Noether potential}
In this subsection we establish the relation between the off-shell ADT potential and Noether potential, and we also establish the relation between off-shell ADT current in the presence of matter fields and the symplectic current in the CPSM.

From Eqs.\eqref{diffv1} and \eqref{diffv2}, and using the off-shell identity \eqref{identity0}, we can obtain the off-shell Noether current
\begin{equation}\label{off-shell Noether current}
J_{\epsilon }^{\mu }=2 \sqrt{-g} \textbf{E}^{\mu \nu } \xi _{\nu }+\sqrt{-g} \xi ^{\mu }  L +\tilde{\Xi }^{\mu }-\tilde{\Theta }^{\mu }\left(  \delta _{\epsilon }\Phi,\Phi \right) ,
\end{equation}
which satisfies $\partial_{\mu} J_{\epsilon }^{\mu }=0$, so that the off-shell Noether potential $K_{\epsilon }^{\mu \nu }$ can be introduced as
\begin{equation}\label{Noether current and potentiali}
J_{\epsilon }^{\mu }=\partial_{\nu} K_{\epsilon }^{\mu \nu } ,
\end{equation}
where $J_{\epsilon }^{\mu }=\sqrt{-g} \mathcal{J}_{\epsilon }^{\mu }, K_{\epsilon }^{\mu \nu }=\sqrt{-g} \mathcal{K}_{\epsilon }^{\mu \nu }$.
The Lie derivative of surface term is
\begin{equation}\label{key}
\mathcal{L}_{\xi }\tilde{\Theta }^{\mu }(\delta \Phi ,\Phi )=\xi ^{\nu } \partial_{\nu} \tilde{\Theta }^{\mu }-\tilde{\Theta }^{\nu } \partial_{\nu} \xi ^{\mu }+\tilde{\Theta }^{\mu } \partial_{\nu} \xi ^{\nu } ,
\end{equation}
which leads to
\begin{equation}\label{ee26}
\xi ^{\mu }\partial_{\nu} \tilde{\Theta }^{\nu }(\delta \Phi ,\Phi )=\partial _{\nu }\left(2 \xi ^{[\mu }\tilde{\Theta }^{\nu ]}(\delta \Phi ,\Phi )\right)+\mathcal{L}_{\xi }\tilde{\Theta }^{\mu }(\delta \Phi ,\Phi ) .
\end{equation}
Since the surface term $\tilde{\Theta }$ is non-covariant,
\begin{align}\label{dsurface}
	\delta _{\epsilon }\tilde{\Theta }^{\mu }(\delta \Phi ,\Phi )&=\delta _{\xi }\tilde{\Theta }^{\mu }(\delta \Phi ,\Phi )+\delta _{\lambda }\tilde{\Theta }^{\mu }(\delta \Phi ,\Phi )
	\nonumber \\&=\mathcal{L}_{\xi }\tilde{\Theta }^{\mu }(\delta \Phi ,\Phi )+\tilde{A}^{\mu }(\delta \Phi ,\delta _{\xi }\Phi,\Phi)+\tilde{\Pi }_{\lambda }^{\mu }(\delta A,A) ,
\end{align}
where $\tilde{A}^{\mu }(\delta \Phi ,\delta _{\xi }\Phi,\Phi)$ and $\tilde{\Pi }_{\lambda }^{\mu }(\delta A,A)$ come from the non-covariance of surface term, and we have denoted $\delta _{\lambda }\tilde{\Theta }^{\mu }(\delta \Phi ,\Phi )=\tilde{\Pi }_{\lambda }^{\mu }(\delta A,A)$.
From Eq.\eqref{dsurface} We obtain
\begin{align}\label{ee28}
		\mathcal{L}_{\xi }\tilde{\Theta }^{\mu }(\delta \Phi ,\Phi )
		=\delta _{\epsilon }\tilde{\Theta }^{\mu }(\delta \Phi ,\Phi )-\tilde{D}^{\mu } ,
\end{align}
where
\begin{align}\label{ee29}
	\tilde{D}^{\mu }&=\tilde{A}^{\mu }(\delta \Phi ,\delta _{\xi }\Phi,\Phi)+\tilde{\Pi }_{\lambda }^{\mu }(\delta A,A)    \nonumber \\&=\tilde{A}^{\mu }(\delta g,\delta _{\xi }g,g)+\tilde{\Pi }_{\lambda }^{\mu }(\delta A,A) ,
\end{align}
here the second equality denotes $\tilde{A}^{\mu }$ term mainly comes from the contribution of gravitational part, which has been considered in \cite{Kim:2013cor} for gravitational Chern-Simons term. For convenience, we consider in our case  an apparently non-covariant term in Lagrangian only consisting of gauge fields, i.e. the matter Chern-Simons term. So the $\tilde{A}^{\mu }$ term vanishes.
By varying the off-shell Noether current \eqref{off-shell Noether current} and using the off-shell ADT current \eqref{off-shell ADT current}, we get
\begin{align}\label{relation}
\delta J_{\epsilon }^{\mu }
=2 \sqrt{-g} \mathcal{J}_{ADT}^{\mu }+\partial _{\nu }\left(2\xi ^{[\mu }\tilde{\Theta }^{\nu ]}(\delta \Phi ,\Phi )\right)-\tilde{\omega }^{\mu }\left(\delta  \Phi ,\delta _{\epsilon }\Phi,\Phi \right)+\delta  \tilde{\Xi }^{\mu }-\tilde{\Pi }_{\lambda }^{\mu } ,
\end{align}
where we have used the Eqs.\eqref{ee26}-\eqref{ee29} and the symplectic current definition \eqref{symplectic current} .

On the other hand, from Eq.\eqref{Noether current and potentiali}, we have
\begin{equation}\label{vij}
\delta J_{\epsilon }^{\mu }=\partial _{\nu }\left(\delta K_{\epsilon }^{\mu \nu }\right) .
\end{equation}
Introducing $\tilde{\Sigma }_{\lambda }^{\mu \nu }$ as
\begin{equation}\label{pi}
\tilde{\Pi} _{\lambda }^{\mu }-\delta \tilde{\Xi} ^{\mu }\equiv \partial _{\nu }\tilde{\Sigma }_{\lambda }^{\mu \nu } ,
\end{equation}
and substituting Eqs.\eqref{vij} and \eqref{pi} into Eq.\eqref{relation}, we obtain
\begin{align}\label{off-shell ADT current and potential2}
	2 \sqrt{-g} \mathcal{J}_{ADT}^{\mu }
	&=\partial _{\nu }\left(\delta K_{\epsilon }^{\mu \nu }-2\xi ^{[\mu }\tilde{\Theta }^{\nu ]}(\delta \Phi ,\Phi )+\tilde{{\Sigma }}_{\lambda } ^{\mu \nu }\right)+\tilde{\omega }^{\mu }\left(\delta  \Phi ,\delta _{\epsilon }\Phi,\Phi \right)    \nonumber  \\& \equiv \partial_{\nu} \left(2 \sqrt{-g} \mathcal{Q}_{ADT}^{\mu \nu }\right) ,
\end{align}
where $\delta _{\epsilon }\Phi  =0$, $\tilde{\omega }^{\mu }\left(\delta \Phi ,\delta _{\epsilon }\Phi,\Phi \right)=0$, for exact symmetry.
 $\mathcal{Q}_{ADT}^{\mu \nu}$ is the off-shell ADT potential corresponding to the off-shell ADT current $\mathcal{J}_{ADT}^{\mu }$, which is given by
\begin{align}\label{off-shell ADT potential}
	2 \sqrt{-g} \mathcal{Q}_{ADT}^{\mu \nu}&=\delta K_{\epsilon }^{\mu \nu }-2\xi ^{[\mu }\tilde{\Theta }^{\nu ]}(\delta \Phi ,\Phi )+\tilde{\Sigma }_{\lambda }^{\mu \nu } .
\end{align}
Now we have established the one-to-one correspondence between the off-shell ADT potential and the Noether potential similar to \cite{Kim:2013zha,Hyun:2014sha}, but in our formalism we have used the exact symmetry $\delta _{\epsilon }\Phi=0$.

In order to obtain finite conserved charges, we use the one parameter path integral method \cite{Wald:1999wa,Barnich:2001jy, Barnich:2003xg, Barnich:2004uw, Barnich:2007bf} in the space of solutions, in which the path is characterized by parameter $s (s\in [0,1])$. This path interpolates between the given solution and the background solution through parameterizing a set of free parameters $\mathcal{M}$ in the space for the solutions of EOM as $s \mathcal{M}$ . Finally by assuming that the integral is path independent, we can define the off-shell ADT conserved charge as
\begin{align}\label{ADT charge}
	   \mathcal{Q}( \epsilon)&\equiv \frac{1}{8 \pi  G}\int _0^1ds\int _{\Sigma }d^{D-2}x_{\mu \nu }\sqrt{-g}\mathcal{Q}_{ADT}^{\mu \nu}    \nonumber  \\&=\frac{1}{16 \pi  G}\int _{\Sigma }d^{D-2}x_{\mu \nu }\left[  \Delta K^{\mu \nu } (\epsilon)-2\xi ^{[\mu }\int _0^1ds\,\,\tilde{\Theta }^{\nu ]}(\Phi ,s\mathcal{M})+\int _0^1ds\,\,\tilde{\Sigma }_{\lambda }^{\mu \nu }(A,s\mathcal{M})\right] ,
\end{align}
where
\begin{equation}\label{key}
\Delta K^{\mu \nu }=K_{s=1}^{\mu \nu }-K_{s=0}^{\mu \nu }
\end{equation}
is the finite difference of Noether potential between the given solution and the background solution. Eq.\eqref{ADT charge} will provide an effective way to compute quasi-local conserved charges. From Eqs.\eqref{off-shell ADT potential} and \eqref{ADT charge},  we can see the elevated off-shell ADT formalism is completely equivalent to the CPSM \cite{Wald:1993nt, Iyer:1994ys, Jacobson:1993vj, Wald:1999wa} and the BBC formalism \cite{Barnich:2001jy, Barnich:2003xg, Barnich:2004uw, Barnich:2007bf}.

\section{A general model in Einstein-Maxwell-Scalar gravity theory with matter Chern-Simons term}
As an application, in this section we calculate the conserved charges for Einstein-Maxwell-Scalar gravity theory with the cosmological constant and a matter Chern-Simons term for any odd dimensions $D=2N+1$. The action has the form
\begin{align}
		I=\frac{1}{16 \pi  G}\int d^{2N+1}x\sqrt{-g}(L_g+L_{\phi}+L_{EM}+L_{CS}) ,
\end{align}
where
\begin{align}
	L_g&=R-2 \Lambda , \nonumber\\
	L_{\phi}&=-\frac{1}{2}f_{AB}(\phi) \partial_ {\mu} \phi ^A\partial ^{\mu }\phi ^B-V(\phi),  \nonumber\\
	L_{EM}&=-\frac{1}{4} \mathcal{N}_{IJ} (\phi)  F_{\mu \nu }^I F^{J\mu \nu },\nonumber \\
	 L_{CS}&=\frac{1}{2}\hat{F}_I^{\mu \nu } F_{\mu \nu }^I ,
\end{align}
and
 \begin{equation}\label{key}
 \hat{F}_I^{\mu \nu }=C_{IJK \cdots L} \epsilon ^{\mu \nu \alpha \beta \gamma \cdots \rho \sigma } A_{\alpha }^J F_{\beta \gamma }^K \cdots F_{\rho \sigma }^L .
 \end{equation}
 Here $\Lambda$ is cosmological constant, and the $\epsilon$-tensor is defined as $\sqrt{-g} \epsilon ^{0 1 2 {\cdots} D-1}=-1$.

 We first consider the covariant part of the theory, i.e. the Einstein-Maxwell-Scalar Lagrangian $L=L_g+L_{\phi}+L_{EM}$. The variation of the Lagrangian is
\begin{equation}\label{key}
\delta  \left(\sqrt{-g} L\right)=\sqrt{-g}\left[\, ^{(g)}E^{\mu \nu }\delta g_{\mu \nu }+\, \, ^{(A)}E_I^{\nu }\delta A_{\nu }^I+\, ^{(\phi) }E_C\delta \phi ^C\right]+\sqrt{-g}\nabla _{\mu }\Theta ^{\mu }(\delta \Phi ,\Phi) ,
\end{equation}
where the Euler-Lagrange expressions are
\begin{flalign}
	\, ^{(g)}E^{\mu \nu }=&-(R^{\mu \nu }-\frac{1}{2}{ g^{\mu \nu }R}+\Lambda g^{\mu \nu })
	 \nonumber \\&+\frac{1}{2}{L_{\phi }g^{\mu \nu } }+\frac{1}{2}f_{A B}(\phi) \nabla ^{\mu }\phi ^A\nabla ^{\nu }\phi ^B
    	\nonumber \\&+\frac{1}{2} \mathcal{N}_{IJ}( \phi) \left[F^{I\mu \alpha } F^{J\nu }_{\, \,\,\,\, \,\,\alpha }-\frac{1}{4}g^{\mu \nu } F_{\alpha \beta }^I F^{J\alpha \beta } \right], \\ \nonumber\\
	\, ^{(\phi)}E_C=&-\frac{1}{2}f_{A B,C}(\phi) \nabla _{\mu }\phi ^A\nabla ^{\mu }\phi ^B-V_{,C}(\phi)
	+\nabla _{\mu }\left(f_{CB}(\phi) \nabla ^{\mu }\phi ^B\right)         \nonumber \\&-\frac{1}{4}  \mathcal{N}_{IJ,C} (\phi) F_{\mu \nu }^I F^{J\mu \nu }, \\ \nonumber\\
	\, ^{(A)}E_I^{\nu }=&\nabla _{\mu }\left[\mathcal{N}_{IJ} (\phi)  F^{J\mu \nu }\right].
\end{flalign}
And the surface terms are given by
\begin{align}
\Theta ^{\mu }(\delta \Phi ,\Phi )&=\Theta _g^{\mu }(\delta g,g)+\Theta _{\phi }^{\mu }(\delta \phi ,\phi )+\Theta _A^{\mu }(\delta A,A)
\nonumber \\&=2\nabla ^{[\sigma }h^{\mu ]}{}_{\sigma }-f_{AB}(\phi) \delta \phi ^A\nabla ^{\mu }\phi ^B-\mathcal{N}_{IJ}(\phi)\delta A_{\nu }^I F^{J\mu \nu },	
\end{align}
where and in what follows
\begin{gather}
	 h_{\mu \nu }=\delta g_{\mu \nu },\,\,\,\,h^{\mu \nu }= g^{\mu \alpha } g^{\nu \beta }\delta g_{\alpha \beta }=-\delta g^{\mu \nu },\,\,\,\,h=g^{\mu \nu }\delta g_{\mu \nu }.
\end{gather}

Take the variation of Lagriangian induced by the generator $\epsilon$
\begin{align}
	\delta _{\epsilon } \left(\sqrt{-g} L\right)&=\sqrt{-g}\left[\,\, ^{(g)}E^{\mu \nu }\delta _{\epsilon }g_{\mu \nu }+\, \, ^{(A)}E_I^{\nu }\delta _{\epsilon }A_{\nu }^I+\, \,^{(\phi )}E_C  \delta _{\epsilon }\phi ^C\right]+\sqrt{-g}\nabla _{\mu }\Theta ^{\mu }\left(\delta _{\epsilon} \Phi ,\Phi \right)
	\nonumber \\&=\sqrt{-g} \nabla _{\mu }\left(\xi ^{\mu } L\right),	
\end{align}
where the transformations are
\begin{align}
	\delta _{\epsilon } g_{\mu \nu }&= \mathcal{L}_{\xi }g_{\mu \nu }=2\nabla _{(\mu }\xi _{\nu )},  \\
	 \delta _{\epsilon }A_{\nu }^I&= \mathcal{L}_{\xi }A_{\nu }^I+\nabla _{\nu }\lambda ^I =\xi ^{\sigma }F_{\sigma \nu }^I +\nabla _{\nu }\left(A_{\sigma }^I\xi ^{\sigma }+ \lambda ^I\right),  \\
	\delta _{\epsilon }\phi ^C &=\mathcal{L}_{\xi }\phi ^C =\xi ^{\mu } \nabla _{\mu }\phi ^C.
\end{align}
The surface term for this transformation is given by
\begin{align}
	\Theta^{\mu}\left(\delta_{\epsilon}\Phi, \Phi \right) =&
	2 \nabla_{\nu} \nabla^{(\mu}\xi^{\nu)} -
	2 \nabla^{\mu} \nabla_{\nu}\xi^{\nu}
	\nonumber\\ &- \mathcal {N}_ {IJ} (\phi) F^{J\mu \nu }\left[\xi^{\sigma} F_ {\sigma\nu}^
	I+ \nabla _ {\nu}\left(A_ {\sigma}^
	I\xi^{\sigma}  + \lambda^I\right) \right]
	\nonumber\\ &-
	f_ {AB} (\phi)\xi^{\nu} \nabla_{\nu}\phi^
	A \nabla^{\mu}\phi^B.
\end{align}
Since we now only consider the covariant part of the theory, the $\Xi ^{\mu }$ term vanishes.

From Eq.\eqref{off-shell Noether current} we obtain the off-shell Noether current and Noether potential as follows
\begin{align}
		 \mathcal{J}_{\epsilon}^{\mu} &=\nabla_{\nu}\left[2 \nabla^{[\mu} \xi^{\nu]}+\mathcal{N}_{IJ}(\phi) F^{J \mu \nu}\left(A_{\sigma}^{I} \xi^{\sigma}+\lambda^{I}\right)\right]
		 \nonumber \\ &=\nabla_{\nu} \mathcal{K}_{\epsilon}^{\mu \nu}, \\
		 \mathcal{K}_{\epsilon}^{\mu \nu} &=2 \nabla^{[\mu} \xi^{\nu]}+\mathcal{N}_{IJ}(\phi) F^{J \mu \nu}\left(A_{\sigma}^{I} \xi^{\sigma}+\lambda^{I}\right). \label{Noether potential2}
\end{align}
From Eq.\eqref{off-shell ADT potential} we have the off-shell ADT potential
\begin{equation}\label{ADT potential2}
\mathcal{Q}_{ADT}^{\mu \nu}(\delta \Phi ,\epsilon)=\mathcal{Q}_{ADT}^{\mu \nu}(\delta g ,\xi)+\mathcal{Q}_{ADT}^{\mu \nu}(\delta \phi ,\xi)+\mathcal{Q}_{ADT}^{\mu \nu}(\delta A ,\epsilon),
\end{equation}
where
\begin{align}
	 \mathcal{Q}_{ADT}^{\mu \nu}(\delta g ,\xi )=&\frac{1}{2} h \nabla^{[\mu} \xi^{\nu]}-h^{\alpha[\mu} \nabla_{\alpha} \xi^{\nu]}-\xi^{[\mu} \nabla_{\alpha} h^{\nu] \alpha}+\xi_{\alpha} \nabla^{[\mu} h^{\nu] \alpha}+\xi^{[\mu} \nabla^{\nu]} h ,  \nonumber \\
	 \mathcal{Q}_{ADT}^{\mu \nu}(\delta \phi ,\xi)=& f_{A B} (\phi) \delta\phi^A \xi ^{[\mu }\nabla ^{\nu ]}\phi^B ,\nonumber \\
	 \mathcal{Q}_{ADT}^{\mu \nu}(\delta A ,\epsilon)=&\frac{1}{2} \biggl[ \big(\frac{1}{2} h \mathcal{N}_{IJ}(\phi) F^{J \mu \nu}+\mathcal{N}_{IJ,C}(\phi) \delta \phi^C F^{J \mu \nu}+\mathcal{N}_{IJ}(\phi) \delta F^{J \mu \nu}\big)  \nonumber \\
	&\cdot \big(A_{\sigma}^{I} \xi^{\sigma}+\lambda^{I}\big) 	
	+\mathcal{N}_{IJ}(\phi) F^{J \mu \nu}\delta A_{\sigma}^{I} \xi^{\sigma}	+2\mathcal{N}_{IJ}(\phi)\xi^{[\mu}F^{J \nu] \sigma}\delta A_{\sigma}^{I} \biggl] .
\end{align}

Different from \cite{Hyun:2014sha,Setare:2016dex}, in our case when we consider the internal gauge transformation, the gauge fields have contributions for off-shell Noether potential \eqref{Noether potential2} and the off-shell ADT potential \eqref{ADT potential2}.

Now we consider the Chern-Simons term $L_{CS}$. The generic variation of the Chern-Simons Lagrangian
\begin{equation}
\delta (\sqrt{-g} L_{C S})=\sqrt{-g}^{(A)} E_{I}^{\alpha} \delta A_{\alpha}^{I}+\sqrt{-g} \nabla_{\mu} \Theta_{C S}^{\mu}(\delta A, A),
\end{equation}
where the Euler-Lagrange expression and the surface term are given by
\begin{align}
	&{^{(A)} E_{I}^{\alpha}=(N+1) \nabla_{\nu} \hat{F}_{I}^{\alpha \nu}}, \\ &{\Theta_{C S}^{\mu}(\delta A, A)=N \hat{F}_{I}^{\mu \nu} \delta A_{\nu}^{I}}.
\end{align}

The variation of the Lagrangian induced by the generator $\epsilon$ has the form
\begin{flalign}
	 \delta_{\epsilon} (\sqrt{-g} L_{C S})=& \sqrt{-g}(N+1) \nabla_{\nu} \hat{F}_{I}^{\alpha \nu}\left[\xi^{\sigma} F_{\sigma \alpha}^{I}+\nabla_{\alpha}\left(A_{\sigma}^{I} \xi^{\sigma}+\lambda^{I}\right)\right] \nonumber \\ &+\sqrt{-g} \nabla_{\mu}\left[N \hat{F}_{I}^{\mu \nu}\left(\xi^{\sigma} F_{\sigma \nu}^{I}+\nabla_{\nu}\left(A_{\sigma}^{I} \xi^{\sigma}+\lambda^{I}\right)\right)\right] .
\end{flalign}
On the other hand,
\begin{align}
	\delta_{\epsilon} (\sqrt{-g} L_{C S})&=\sqrt{-g} \nabla_{\mu}\left[\xi^{\mu} L_{CS}+\Xi ^{\mu }\right] ,
\end{align}
where
 \begin{equation}\label{key}
 \Xi ^{\mu }=\lambda^{I} \nabla_{\nu } \hat{F}_{I}^{\mu \nu}
 \end{equation}
comes from the non-covariance of the matter Chern-Simons Lagrangian.

By similar procedure to the covariant part, we can obtain off-shell Noether current and potential
\begin{align}
	\mathcal{J}_{CS}^{\mu}&=\nabla_{\nu}\left[-N \hat{F}_{I}^{\mu \nu}\left(A_{\sigma}^{I} \xi^{\sigma}+\lambda^{I}\right)\right], \\ \mathcal{K}_{CS}^{\mu \nu}&=-N \hat{F}_{I}^{\mu \nu}\left(A_{\sigma}^{I} \xi^{\sigma}+\lambda^{I}\right).
\end{align}
From Eq.\eqref{pi}, the additional terms are
\begin{align}
	 \Pi_{\lambda}^{\mu}(\delta A, A) &=N \delta _{\lambda ^I} \hat{F}_I^{\mu \nu } \delta A_{\nu }^I ,\\ \Sigma_{\lambda}^{\mu \nu}(\delta A, A) &=-N C_{I J K \cdots L} \epsilon^{\mu \nu \alpha \beta \gamma \cdots \rho \sigma} \lambda^{I} \delta A_{\alpha}^{J} F_{\beta \gamma}^{K} \cdots F_{\rho \sigma}^{L}.
\end{align}
From Eq.\eqref{off-shell ADT potential}, The contribution of the Chern-Simons term  for the off-shell ADT potential is given by
\begin{equation}
2 \sqrt{-g} \mathcal{Q}_{ADT}^{\mu \nu}=\delta\left(\sqrt{-g} \mathcal{K}_{CS}^{\mu \nu }\right)-2 \sqrt{-g} \xi^{[\mu} \Theta_{CS}^{\nu]}(\delta A, A)+\sqrt{-g}\Sigma_{\lambda}^{\mu \nu }(\delta A, A).
\end{equation}
The formulas above have already been calculated for Chern-Simons theory in \cite{Compere:2009dp, Barnich:2005kq,Rogatko:2007pv,Suryanarayana:2007rk,Hanaki:2007mb} in CPSM and the BBC formalism up to a supplementary term, which does not affect the final integral results for conserved charge calculation. Therefore, we have explicitly shown the equivalence of our formulation with the CPSM and the BBC formalism.

\section{Conserved charges of general  G\"odel black holes in minimal five-dimensional gauged supergravity}

When $D=5, \Lambda=0,$ and $D=3, \Lambda \neq 0$ \footnote{The cosmological constant does not affect the conserved current and potential if it is fixed, but it affects the black hole solutions (for example AdS ), so it affects the final conserved charges.}, we have checked that our off-shell ADT potentials are completely equivalent to BBC surface charges obtained by Comp\`ere and collaborators in \cite{Barnich:2005kq} and \cite{Banados:2005da}, respectively. By calculation we can get the same results of mass, angular momentum and electric charge of G\"odel black holes as those they got.

As an explicit example, in this section we use the elevated off-shell ADT formulation to calculate the conserved charges of general non-extremal rotating charged G\"odel black holes in minimal five-dimensional gauged supergravity. This general Einstein-Maxwell-Chern-Simons-G\"odel (EMCS-G\"odel) black hole was obtained by Wu in \cite{Wu:2007gg} as
\begin{align}
			d s^{2}=&-f(r) d t^{2}-2 g(r) \sigma_{3} d t+h(r) \sigma_{3}^{2}
			+\frac{d r^{2}}{V(r)}+\frac{r^{2}}{4} d \Omega_{3}^{2}  \label{line element}, \\
			A=& B(r) d t+C(r) \sigma_{3}, \label{potential}
\end{align}
where the unit $3$-sphere $d \Omega_{3}^{2}$ and the left invariant form $\sigma_{3}$ are given by
\begin{gather}
	\qquad d \Omega_{3}^{2}=d \theta^{2}+\sin ^{2} \theta d \psi^{2}+\sigma_{3}^{2}, \quad \sigma_{3}=d \phi+\cos \theta d \psi ,
\end{gather}
and
\begin{align*}
	f(r) =&1-\frac{2 m}{r^{2}}+\frac{q^{2}}{r^{4}} \\ g(r) =&j r^{2}+3 j q+\frac{(2 m-q) a}{r^{2}}-\frac{q^{2} a}{2 r^{4}} \\ h(r) =&-j^{2} r^{2}\left(r^{2}+2 m+6 q\right)+3 j q a \\ &+\frac{(m-q) a^{2}}{2 r^{2}}-\frac{q^{2} a^{2}}{4 r^{4}} \\ V(r) =&1-\frac{2 m}{r^{2}}+\frac{8 j(m+q)[a+2 j(m+2 q)]}{r^{2}} \\ &+\frac{2(m-q) a^{2}+q^{2}\left[1-16 j a-8 j^{2}(m+3 q)\right]}{r^{4}} \\
	B(r)=&\frac{\sqrt{3} q}{2 r^{2}} \\\quad C(r)=&\frac{\sqrt{3}}{2}\left(j r^{2}+2 j q-\frac{q a}{2 r^{2}}\right)
\end{align*}
is a solution to the theory
\begin{align}
	 L &=L_{gEM}+L_{CS}  \nonumber \\ &=R-F_{\mu \nu } F^{\mu \nu}-\frac{2}{3 \sqrt{3}} \epsilon^{\gamma \alpha \beta  \mu \nu } A_{\gamma} F_{\alpha \beta} F_{\mu \nu}.
\end{align}

From our previous conserved current and potential, when we take $N=2,\,\mathcal{N}_{IJ} (\phi)=4,\,C_{IJK \cdots L}=- \frac{4}{3 \sqrt3}$ and drop the $IJK\cdots L$ indices the relevant quantities can be summarized as
\begin{align}
	&\Theta_{gEM}^{\mu}(\delta g, g ; \delta A, A)=2 \nabla^{[\sigma} {h^{\mu]}}_{\sigma}-4 \delta A_{\nu} F^{\mu \nu}, \nonumber \\
	&\mathcal{K}_{gEM}^{\mu \nu}(g,A,\epsilon)=2 \nabla^{[\mu} \xi^{\nu]}+4 F^{\mu \nu}\left(A_{\sigma} \xi^{\sigma}+\lambda\right), \nonumber \\
	&\Theta_{CS}^{\mu}(\delta A, A)=- \frac{8}{3 \sqrt{3}} \epsilon^{\mu \nu \alpha \beta \gamma} A_{\alpha} F_{\beta \gamma} \delta A_{\nu},   \nonumber \\
	&\mathcal{K}_{CS}^{\mu \nu}(A, \epsilon)= \frac{8}{3 \sqrt{3}} \epsilon^{\mu \nu \alpha \beta \gamma} A_{\alpha} F_{\beta \gamma}\left(A_{\sigma} \xi^{\sigma}+\lambda\right),  \nonumber \\
	&\Sigma_{\lambda}^{\mu \nu}(\delta A, A)= \frac{8}{3 \sqrt{3}} \epsilon^{\mu \nu \alpha \beta \gamma} \delta A_{\alpha} F_{\beta \gamma} \lambda .
\end{align}

In order to obtain finite conserved charges, we use one parameter path integral method \cite{Wald:1999wa,Barnich:2001jy, Barnich:2003xg, Barnich:2004uw, Barnich:2007bf} to calculate the mass, angular momentum and electric charge of EMCS-G\"odel black hole. The path $\gamma :\Phi(s\mathcal{M})$ interpolating between the background G\"odel-type universe $\bar{\Phi }$ and the EMCS-G\"odel black hole solution $\Phi$ is obtained by substituting $(m,a,q)$ by $(sm,sa,sq)$ in \eqref{line element} and \eqref{potential}, with $s\in [0,1]$. Then the integration is performed along that path as $\mathcal{Q} (\epsilon) =\int _{0}^1ds\,\,\delta \mathcal{Q} (\epsilon ;s \mathcal{M})$. For convenience, we choose surface $\mathcal{S}:t=const=r$ to integrate. From Eq.\eqref{ADT charge}, the conserved charge has the form
\begin{align}
					  \mathcal{Q}(\epsilon)=\frac{1}{16 \pi  G}\int _{\Sigma }d^{D-2}x_{\mu \nu } \biggl[\Delta K_{gEM}^{\mu \nu }&-2\xi ^{[\mu }\int _0^1ds\,\, \tilde{\Theta}_{gEM}^{\nu ]}(\Phi ,s\mathcal{M})+\Delta K_{CS}^{\mu \nu }   \nonumber \\  &-2\xi ^{[\mu }\int _0^1ds\,\, \tilde{\Theta} _{CS}^{\nu ]}(\Phi ,s\mathcal{M})+\int _0^1ds\,\, \tilde{\Sigma} _{\lambda }^{\mu \nu }(A,s\mathcal{M})\biggl].
\end{align}

 The time translational timelike Killing vector is taken as $\xi =\xi ^{\mu }\partial _{\mu },\,\xi ^{\mu }=(-1,0,0,0,0)$, and the generator $\epsilon =(\xi ,0)$, which satisfies the exact symmetry $\delta _{\epsilon }\Phi=0$. The Noether potentials and the surface terms are calculated as
\begin{align*}
	\Delta K_{gEM}^{ t r}=&\frac{1}{2} \left(-8 j^2 m^2-24 j^2 m q-20 j^2 q^2+m\right)\sin \theta-2 j^2 m r^2 \sin \theta   \\&+\frac{q^2  \left(8 a j-8 j^2 (m+3 q)+1\right)\sin \theta}{4 r^2}-\frac{a q^2 (a-12 j q) \sin \theta}{4 r^4}-\frac{a^2 q^3 \sin \theta }{2 r^6},  \\
	\int _0^1ds\,\, \tilde{\Theta} _{gEM}^{r}=&-\frac{1}{4}  \left(m \left(4 a j+24 j^2 q-1\right)+4 j q (a+6 j q)+16 j^2 m^2\right)\sin \theta+2 j^2 m r^2 \sin \theta     \\ &+\frac{q^2  \left(8 j^2 (m-q)-1\right)\sin \theta}{4 r^2}+\frac{a q^2 (a+6 j q)\sin \theta}{4 r^4}-\frac{a^2 q^3 \sin \theta }{10 r^6} ,  \\
	\Delta K_{CS}^{t r}=&8 j^2 q^2 \sin \theta +2 j^2 q r^2 \sin \theta+\frac{2 j q^2  (4 j q-a)\sin \theta}{r^2} -\frac{4 a j q^3 \sin \theta }{r^4}+\frac{a^2 q^3 \sin \theta }{2 r^6},   \\
	\int _0^1ds\,\, \tilde{\Theta} _{CS}^{r }=&-2 j^2 q^2 \sin \theta -2 j^2 q r^2 \sin \theta -\frac{a j q^3 \sin \theta }{2 r^4} +\frac{a^2 q^3 \sin \theta }{10 r^6} ,  \\
\end{align*}
from which the mass of EMCS-G\"odel black hole is given by
\begin{align}\label{totall mass}
	M =&\frac{1}{16 \pi  G}\int _{\Sigma }d^3x_{t r} \biggl[\Delta K_{gEM}^{ t r}-2\xi ^{[t }\int _0^1ds\,\, \tilde{\Theta} _{gEM}^{r]}(\Phi ,s\mathcal{M})+\Delta K_{CS}^{t r}    \nonumber \\  &-2\xi ^{[t }\int _0^1ds\,\, \tilde{\Theta} _{CS}^{r ]}(\Phi ,s\mathcal{M})\biggl]   \nonumber \\
	=& \frac{\pi}{G}\left[\frac{3}{4} m-j(m+q) a-2 j^{2}(m+q)(4 m+5 q)\right].	
\end{align}
We can see that the mass not only comes from the gravitational part, but the electromagnetic and matter Chern-Simons term have also contributions.

Taking the rotational Killing vector $\xi =\xi ^{\mu }\partial _{\mu },\,\xi ^{\mu }=(0,0,0,1,0),\epsilon =(\xi ,0)$. Since $\xi$ is tangent to the integral surface, the surface terms have no contributions to the angular momentum. The Noether potentials are calculated as
\begin{align*}
	\Delta K_{gEM}^{ t r}=&-\frac{1}{4} \big(4 a^2 j (m-q)+a \left(16 j^2 m^2+16 j^2 m q+16 j^2 q^2-2 m+q\right)   \\ &-6 j q^2 \left(8 j^2 (m+3 q)-1\right)\big) \sin \theta -3 j^2 q r^2   (a-8 j q)\sin \theta+12 j^3 q r^4 \sin \theta     \\ &+\frac{3 a j q^2 (a-8 j q)\sin \theta}{2 r^2}+\frac{3 a^2 j q^3 \sin \theta }{r^4}-\frac{a^3 q^3 \sin \theta }{4 r^6}, \\
	\Delta K_{CS}^{t r}=&12 a j^2 q^2 \sin \theta-16 j^3 q^3 \sin \theta +3 j^2 q r^2 (a-8 j q)\sin \theta -12 j^3 q r^4 \sin \theta     \\ &-\frac{3 a j q^2 (a-8 j q)\sin \theta}{2 r^2}-\frac{3 a^2 j q^3 \sin \theta }{r^4}+\frac{a^3 q^3 \sin \theta }{4 r^6},
\end{align*}
then we obtain the angular momentum
\begin{align}
	J_{\phi}=&\frac{1}{16 \pi  G}\int _{\Sigma }d^3x_{t r} \biggl[\Delta K_{gEM}^{ t r}+\Delta K_{CS}^{t r}\biggl]  \nonumber \\
	=&\frac{\pi}{2 G}  \biggl[a\left(m-\frac{q}{2}-2 j(m-q) a-8 j^{2}\left(m^{2}+m q-2 q^{2}\right)\right) \nonumber \\ &-3 j q^{2}+8 j^{3}(3 m+5 q) q^{2}\biggl] ,
\end{align}
while the angular momenta for other rotational killing vectors vanish.

When $\epsilon =(0 ,-1)$, which is a solution of $\delta _{\epsilon }\Phi=0$, the additional non-covariant term has a contribution to electric charge. The Noether potentials and the additional terms are calculated as
\begin{align*}
	\Delta K_{gEM}^{ t r}=&-\frac{1}{2} \sqrt{3} \left(4 a j m+q \left(8 j^2 (m+3 q)-1\right)\right) \sin \theta-8 \sqrt{3} j^2 q r^2 \sin \theta \\& +\frac{4 \sqrt{3} a j q^2 \sin \theta }{r^2} -\frac{\sqrt{3} a^2 q^2 \sin \theta }{2 r^4},   \\
	\Delta K_{CS}^{t r}=&\frac{4 j q (-a+4 j q )\sin \theta}{\sqrt{3}} +\frac{16 j^2 q r^2 \sin \theta }{\sqrt{3}}-\frac{8 a j q^2 \sin \theta }{\sqrt{3} r^2} +\frac{a^2 q^2 \sin \theta }{\sqrt{3} r^4},  \\
	\int _0^1ds\,\, \tilde{\Sigma} _{\lambda }^{t r }=&\frac{2 j q (-a+4 j q )\sin \theta}{\sqrt{3}} +\frac{8 j^2 q r^2 \sin \theta }{\sqrt{3}}-\frac{4 a j q^2 \sin \theta }{\sqrt{3} r^2}+\frac{a^2 q^2 \sin \theta }{2 \sqrt{3} r^4} ,  \\
\end{align*}
from which the electric charge is given by
\begin{align}
	Q &=\frac{1}{16 \pi  G}\int _{\Sigma }d^3x_{t r } \biggl[\Delta K_{gEM}^{t r }+\Delta K_{CS}^{t r }+\int _0^1ds\,\, \tilde{\Sigma} _{\lambda }^{t r }(A,s\mathcal{M})\biggl]  \nonumber \\
	&=\frac{\sqrt{3} \pi}{2 G} \left[q-4 j(m+q) a-8 j^{2}(m+q) q\right].
\end{align}

Finally, if we take the generator $\epsilon=(-\partial_{t}-\Omega^{\phi}_H \partial_{\phi},\,\Phi_H) $, analogous to the CPSM \cite{Wald:1993nt,Iyer:1994ys}, we can define black hole entropy as the conserved charge
\begin{equation}
\frac{\kappa }{2 \pi }\delta S=\frac{1}{8 \pi  G}\int _{\mathcal{H}}d^{D-2}x_{\mu \nu }\sqrt{-g}\mathcal{Q}_{ADT}^{\mu \nu} .
\end{equation}
If we consider the linearity of the off-shell ADT potential with respect to generator $\epsilon$, it is easy to verify the first law of black hole thermodynamics
\begin{equation}\label{The first law}
\delta M=T_H\delta S + \Omega _H^{\phi }\delta J_{\phi }+\Phi _H\delta Q  .
\end{equation}
The first law can also be checked by substituting the entropy \cite{Wu:2007gg} $S={\mathcal{A}}/{4}$ and the Hawking temperature $T_H={\kappa }/{2 \pi }$, where $\mathcal{A}$ is the horizon area of black hole and $\kappa$ is the surface gravity.

In the above we have chosen the minus sign for angular momentum and to satisfy the first law of black hole thermodynamics. In expression \eqref{The first law} we have considered the G\"odel parameter  $j$ as a fixed constant.
The above conserved charges are completely agree with the ones in \cite{Wu:2007gg}. Therefore, we have explicitly shown the validity of our elevated off-shell ADT conserved charge formalism in the presence of arbitrary matter fields.

\section{Conclusion and Discussions}

In this paper we elevated the formalism for off-shell ADT conserved charges in the presence of arbitrary matter fields to including the internal gauge transformation, when the gauge fields are present. For this purpose, we resort to exact symmetry, and the symmetry generator is combined by diffeomorphism and gauge transformation. In this procedure we are not restricted to Killing vector for diffeomorphism. For universality, we consider an apparently non-covariant Lagrangian, which contains a matter Chern-Simons term. The elevated off-shell ADT formalism provides an efficient way to compute the quasi-local conserved charges in the presence of gauge field. We have shown that the elevated off-shell ADT formalism is completely equivalent to the CPSM and the BBC formalism. Finally, we computed the conserved charges of general non-extremal rotating charged G\"odel black holes in minimal five-dimensional gauged supergravity, which agree with the previous known results. The off-shell ADT formulation can be used even with slow falloff matter fields in asymptotic infinity \cite{Gim:2014nba}. It also does not need to add counter terms to the Lagrangian \cite{Stelea:2008tt} in order to get the correct results for our formalism.

As a comparison, in the traditional linearized ADT formalism, the ADT potential is obtained by linearizing the EOM with respect to a fixed background vacuum metric, which is usually asymptotically flat or (A)dS, and the ADT charge merely involves the gravitational field and its perturbation. Thus, the traditional linearized ADT formalism is based on the assumptions of the fast falloff of matter fields and the perturbed metric at the infinity. The matter fields finally make no contribution to the conserved charges. So the original ADT formulation may succeed to produce the physical conserved charges when the matter fields fall off fast enough at infinity. But usually the spacetime is not asymptotically flat or (A)dS and the matter fields are with slow falloff condition, in which the contributions from the matter fields must be considered and the traditional linearized ADT method is invalid. As a non-trivial generalization of the traditional linearized ADT method, the off-shell ADT method does not resort to the background vacuum metric and the asymptotic behavior of spacetime, i.e., it is valid for any spacetime with any asymptotic structure including asymptotic (A)dS and asymptotic Lifshitz. Our formalism is the elevation of the off-shell method in the presence of arbitrary matter fields to including the internal gauge transformation. The effect of matter fields is contained in our formalism, as in Eq.\eqref{totall mass} the electromagnetic field has a contribution to the mass of the EMCS-G\"odel black hole. In our off-shell ADT formalism it is not necessary to consider concretely the falloff condition of the matter fields and the asymptotic structure of the spacetime. For slowly decaying matter fields that may yield divergent contributions to the conserved charges, by using the one parameter path integral method the conserved charges are automatically regular and need not to be renormalized. At this point our formalism has the same advantages as the SPSM and the BBC formalisms.

Usually, the effect of matter fields is very important for generic theories of gravity. The off-shell ADT formalism will provide a systematic way to construct conserved quantities. In the future work we will generalize the off-shell ADT formalism to including the non-Abelian symmetry, when the non-Abelian gauge fields are present. There exists an extensive class for supersymmetry black hole solutions in the presence of non-Abelian gauge field \cite{Ortin:2015hya}. It will be an intrigue and important issue to study the conserved quantities for this class of solutions.

On the other hand, as we mentioned in the INTRODUCTION, there are various methods proposed to compute the conserved charges in gravity theories, each having their own merits and demerits. Moreover, a given black hole solution might have quite different conserved charges by using different method or in different gravity theories. For example, for three-dimensional Oliva-Tempo-Troncoso (OTT) black hole solution \cite{Oliva:2009ip} in BHT or `New Massive Gravity' (NMG), four-dimensional black hole in Weyl gravity \cite{Lu:2012xu}, three- and five-dimensional Lifshitz black holes in quadratic curvature gravity \cite{Devecioglu:2010sf}, and G\"odel black hole, etc., the traditional ADT method cannot give the correct conserved charges consistent with the first law of black hole thermodynamics. When the generator of gauge transformation $\lambda$ is set to zero and the diffeomorphism generator $\xi$ is taken as killing vector filed our formalism reduces to the cases in \cite{Kim:2013zha, Hyun:2014sha}. In \cite{Hyun:2014sha} the equivalence of the off-shell ADT formalism with the boundary stress tensor method \cite{Balasubramanian:1999re} for the asymptotic AdS spacetime to obtain holographic conserved charges consistent with the dual field theory has been proved. The off-shell ADT formalism already has many successful applications \cite{Gim:2014nba,Hyun:2016dvt,Kulkarni:2019nrk,Peng:2014gha,Peng:2015yjx,Peng:2016wzr,Adami:2017phg} when the diffeomorphism is restricted to Killing vector field. By using the off-shell ADT method the problem presented in \cite{Lu:2012xu} has been overcome in \cite{Peng:2014gha} in Weyl and Einstein-Gauss-Bonnet gravities and the correct conserved charges are obtained. The problem presented in \cite{Devecioglu:2010sf} has been overcome in \cite{Gim:2014nba} for three- and five-dimensional Lifshitz black holes by using this method and the results are confirmed by Padmanabhan method, in which the conserved charges are consistent with the first law of black hole thermodynamics. Furthermore, the quasilocal conserved charges have been checked in topologically massive gravity (TMG) in \cite{Kim:2013cor}. For OTT black hole, by using the off-shell ADT method in the first order gravity formalism, the conserved charges have been examined in review article \cite{Adami:2017phg} (see Eq. (499) and (518) therein), in which the mass and entropy are consistent with the first law of black hole thermodynamics but they are both twice the ones obtained from GR when the OTT black hole reduced to the BTZ one. The same results are obtained in \cite{Nam:2016pfp} by using the Wald formalism in the first order gravity formalism, and confirmed in \cite{Nam:2010dd}. Actually, in this paper we proved the equivalence of the off-shell ADT method with the SPSM and the BBC formalisms, in which many conserved charges, such as the conserved charges in three-dimensional Lifshitz black hole and in warped $\mathrm{AdS_3}$ black hole in NMG \cite{Ghodrati:2016vvf}, have been checked to have the correct values consistent with the first law of black hole thermodynamics.

Although the off-shell ADT formalism already has many successful applications, there are still quite a few problems on the conserved charges to be studied. For example, the conserved charges for OTT black hole solution are obtained by off-shell ADT method and CPSM in \cite{Adami:2017phg} and in \cite{Nam:2016pfp},respectively. Though the results coincide and consistent with the first law of black hole thermodynamics, the values of the conserved charges are twice of the ones obtained from GR when the OTT black hole reduces to the static and neutral BTZ black hole. For the $ \mathrm{Warped-AdS_3}$ black hole with a scalar field \cite{Giribet:2015lfa} as mentioned in \cite{Peng:2015yjx}, the mass and angular momentum are zero by using the off-shell ADT method. These problems are important directions for further study on the conserved charges.

\section*{Acknowledgement}%
\label{sec:acknowledgement}
This work is supported partially by the National Science Foundation of China under Grants No. 11847080
and No. 10671128
and by the Key Project of Chinese Ministry of Education under Grant No. 211059.



\end{document}